\documentstyle[seceq,epsf,twoside]{ptptex}
\notypesetlogo  
\title{ $\tilde U(12)$ A New Symmetry\\ Possibly Realizing in
 Hadron Spectroscopy}
\author{%
Shin {\sc Ishida} and Muneyuki {\sc Ishida}$^{*}$  }
\inst{%
Atomic Energy Research Institute, 
College of Science and Technology\\
Nihon University, Tokyo 101-0062, Japan\\
$^*$ Department of Physics, Tokyo Institute of Technology\\
Tokyo 152-8551, Japan
}
\recdate{%
\today
}
\abst{%
Starting from the multi-local Klein-Gordon equations with Lorentz-scalar 
squared-mass operator we give a covariant quark representation of the 
general composite mesons and baryons with definite Lorentz transformation 
property. The phenomenologically observed hadron mass spectra is pointed
out to satisfy possibly the approximate symmetry under the $\tilde U(4)$ 
transformation group concerning the spinor freedom of light constituent 
quarks, including the chiral transformation as a subgroup. This symmetry 
predicts the existence of new type of chiral mesons and baryons out of 
the conventional framework in non-relativistic quark model:
For light $q\bar q$ systems, the scalar $\sigma$- and 
axial-vector $a_1$-nonets are predicted to exist as relativistic
$S$-wave states besides the ordinary $P$-wave state mesons. 
Two ``exotic" $J^{PC}=1^{-+}$ mesons are predicted to exist as relativistic $P$-wave states,
which are possibly assigned as $\pi_1(1400)$ and $\pi_1(1600)$.
For light 
quark baryons the extra \underline{\bf 56} with positive parity and the 
extra \underline{\bf 70} with negative parity of the static $SU(6)$ are 
predicted to exist as the ground state chiral particles.
}

\begin{document}
\maketitle

\setcounter{tocdepth}{4}

\section{Introduction}
\hspace*{-0.8cm}(\underline{\it The\ present\ status\ of\ 
level-classification\ of\ hadrons})\ \ \ \ 
There exist the two contrasting, non-relativistic and relativistic, 
viewpoints of level-classification. The former is based on the 
non-relativistic quark model (NRQM) with the approximate $LS$-symmetry 
and gives a theoretical base to the PDG level-classification. The 
latter is embodied typically in the NJL model with the approximate 
chiral symmetry. It is widely accepted that $\pi$ meson nonet has 
the property as a Nambu-Goldstone boson in the case of spontaneous 
breaking of chiral symmetry.

\begin{table}
\begin{center}
\begin{tabular}{l|c|c}
\hline
 & {\bf Non-Relativistic} & {\bf Relativistic} \\
\hline
 Model & NRQM & NJL model \\
\hline
 Approx. Symm. & $LS$-Symm. & Chiral Symm. \\
\hline
 Evidence & Bases for PDG & $\pi$ nonet as NG boson\ \ \ \ \ \ \ \   \\
\hline
\end{tabular}
\end{center}
\caption{Two Contrasting Viewpoints of Level Classification}
\label{tab1}
\end{table}

Owing to the recent progress, both theoretical and experimental,
the existence of light $\sigma$-meson as chiral partner of $\pi (140)$
seems to be established\cite{rf1} especially through the analysis of 
various $\pi\pi$-production processes. This gives further a strong 
support to the relativistic viewpoint.

Thus, the hadron spectroscopy is now confronting with a serious problem,
existence of the seemingly contradictory two viewpoints, Non-relativistic 
and Extremely Relativistic ones.\\
(\underline{\it The\ purpose\ of\ this\ letter})\  is, unifying these 
two viewpoints, to propose a covariant level-classification scheme of 
light-through-heavy quark (and possibly of gluonic) hadrons. We shall 
point out a possibility that an approximate symmetry of 
$\tilde U(12) \supset \tilde U(4)_{\rm D.S.} \bigotimes U(3)_{\rm F}$,
here $\tilde U(4)_{\rm D.S.}$ (we denote the homogeneous Lorentz group, 
${\cal L}_4$, as $\tilde U(4)_{\rm D.S.}$) $(U(3)_{\rm F})$ being a 
pseudo-unitary Lorentz (unitary) group concerning Dirac spinor (flavor) 
of light quarks, is realized in nature in the world of hadron spectroscopy.

Here the $\tilde U(12)$ symmetry is mathematically the same as the one 
that appeared in 1965 to generalize covariantly the static $SU(6)_{\rm SF}$ 
symmetry ($SU(6)_{\rm SF} \supset SU(2)_{\rm S}\bigotimes U(3)_{\rm F}$)
in NRQM. However, in the case of $\tilde U(12)_{\rm SF}$ at that time 
\underline{only\ the} \underline{boosted\ Pauli-spinors} are taken as 
\underline{physical components} out of the fundamental representations of 
$\tilde U(4)_{\rm D.S.}$. Now in the present scheme of the $\tilde U(12)$ 
symmetry \underline{all\ general} \underline{Dirac\ spinors} are, inside of hadrons, 
to be treated as \underline{physical}.\\
(\underline{\it History\ of\ symmetry\ and\ level\ classification})\ \ \ \   
First we shall review briefly a history of the hadron-level
classification since the birth of hadron physics. It is 
based on a representation of some symmetry, which 
is deeply connected to the composite picture of hadrons.

In 1933 Yukawa\cite{Yukawa35} predicted existence of $\pi$-meson,
``symbolic particle of Strong Interaction and Hadron Physics."

In 1949 Fermi and Yang\cite{FY} had proposed the composite model 
of $\pi$-mesons. In 1956 Sakata\cite{Sakata56}, extending the 
framework of F.Y. model, proposed a composite model for the new 
particles (mesons and baryons), taking Sakata triplet $S(P,N,\Lambda )$
as the fundamental constituents. 
The strangeness of Nishijima-Gell-Mann rule\cite{NGM} was
identified with the number of $\Lambda$-particle. 
In 1959 Ikeda, Ogawa and Ohnuki\cite{IOO} presented 
the mathematical framework of $U(3)$ symmetry in Sakata model, 
after an important notice by Ogawa\cite{Ogawa} that 
each member of Sakata triplet has a certain equality, 
considered from the role of its numbers $N_i$ $(i=P,N,\Lambda )$ 
as quantum numbers, suggesting a possible certain symmetry.
The similar approach was proposed by Yamaguchi, too 

The choice of Sakata triplet as a fundamental triplet was inadequate,
leading to a difficulty in baryon assignment, and replaced\cite{Sakata63} 
with the (light) quark triplet $q(u,d,s)$ by Gell-Mann and Zweig\cite{GZ} 
in 1964. Shortly after the appearance of quark model in 1964, Sakita\cite{B5}, 
G\"ursey-Radicati\cite{B4} and Pais\cite{Pais} independently proposed the 
$SU(6)_{\rm SF}$ theory, treating the intrinsic spin of quarks as a 
group-theoretical object.

The success of $SU(6)$ theory was remarkable in that 
i) all the ground state mesons and baryons experimentally observed 
   are assigned satisfactorily to its $\underline{\bf 35}$ and 
   $\underline{\bf 56}$ representation, respectively.
ii) Moreover, the famous ratio of nucleon magnetic moments 
   $\mu_P/\mu_N=-3/2$, is predicted.
iii) Furthermore, concerning the $P$-wave Yukawa interaction of mesons 
   with baryons, the desirable $F/D$ ratio is insisted\footnote{
   However, note that this interaction has the zero static limit.}
   to be also obtained.

However, all the above results are depending on the assignment of baryons 
into the corresponding symmetrical  $\underline{\bf 56}$ representation, 
which seems to be, in simple intuition, contradictory with the sacred 
spin-statistics connection of elementary particles.

In 1965 Salam, Delbourgo and Strathdee\cite{SDS}, and Sakita-Wali\cite{SW}
independently proposed the $\tilde U(12)$ symmetry as a relativistic 
extension of the $SU(6)_{\rm SF}$ symmetry, and set up the Bargmann-Wigner 
Equation to be satisfied by free hadron wave functions. Then assigning the 
baryons and mesons to the completely symmetrical representation 
$\underline{\bf 364}$ and $\underline{\bf 144}$, respectively, Salam et al. 
showed that all the above desirable results of the $SU(6)$ (, including the 
$F/D$ ratio of Yukawa interaction) is reproduced. However, they assumed 
there that the ``boosted multi-Pauli spinors", which reduce to the 
multi-Pauli spinors at hadron rest frame, are only allowed to be physical 
and the effective vertex is to be (``fundamentally broken")\footnote{
Note that the Bargmann-Wigner equation with a definite mass itself is 
\underline{not} $\tilde U_{\rm DS}(4)$ (accordingly $\tilde U(12)$)-symmetric. 
} 
$\tilde U(12)$ symmetric. It is now well-known that a relativistic extension
of the $SU(6)_{\rm SF}$ symmetry as a mathematical group is impossible 
as no-go theorem\cite{CM}.  

In 1968 Roman and one of the authors(S.I.)\cite{IR} derived the SU(6) 
results from a different physical consideration on its covariant 
generalization, applying a kind of the bosonization method. That is, 
starting from general Lorentz invariant 4 quark interactions and 
imposing them a ``rest-condition," to be $SU(6)$ symmetric at the 
rest frame of all quarks in a channel where all quark and anti-quark 
numbers are separately conserved.

In order to treat the excited hadrons it is necessary to introduce 
the freedom of relative space-coordinates between constituent quarks. 
This is a straight-forward extension of the SU(6) theory from the 
composite picture, which was developed (first proposed by 
Greenberg\cite{Greenberg64}, 1964) by many authors in the (so-called, 
simple and realistic) symmetrical quark model. They assumed there 
that constituent quarks, inside of hadrons, behave 
non-relativistically\cite{Morpurgo65}, leading to the approximate 
$LS$-symmetry\cite{ID} in the hadron spectroscopy. So the hadron 
wave functions in this scheme is the tensors in the 
$SU(6)_{\rm SF}\bigotimes O(3)_{\rm L}$ space.

In 1970 one of the present authors (S.I.) proposed its relativistic 
generalization, the Urciton Scheme\cite{Ishida70} (, applying 
exciton-picture\footnote{
Here the strange symmetric properties (, now accepted to be as the 
color-singlet behavior) of baryon wave function is reduced to a 
bose-quantization of (ur-)exciton quarks\cite{HG}.
} 
of constituent quarks), where the kinematical region is extended 
into Boosted $SU(6)_{\rm SF}\bigotimes O(3,1)$ space ($O(3,1)$ being Lorentz space),
and accordingly the framework is manifestly covariant. This scheme gives an effective
method to treat both hadron spectroscopy and reactions on common footing
and has been applied actually for these three decades 
as the covariant oscillator quark model\cite{COQM}.

\section{Covariant Framework for Describing Composite Hadrons}
\hspace*{-0.8cm}(\underline{\it Wave\ Functions\ of\ Mesons\ and\ Baryons})
\ \ \ \  In order to introduce the quark-composite picture of hadrons,
we set up the following wave functions(WF) for mesons and baryons, 
respectively.
\begin{eqnarray}
{\rm Meson} &:& \Phi_A{}^B(x_1,x_2)\ ,\ \ \ \   
{\rm Baryon}\ \  : \ \  \Phi_{A_1 A_2 A_3}(x_1,x_2,x_3)\ ; 
\label{eq2}
\end{eqnarray}
where $A,\ A_i(\alpha , a);\ \alpha =1\sim 4 (\ a\ )$ denote Dirac(flavor)
indices of respective quark constituents. $B=(\beta ,b)$ its conjugate
ones; and $x_1$ etc. denote the Lorentz four-vectors of the space-time 
coordinates of constituents. \\
(\underline{\it Klein-Gordon\ equation\ and\ mass\ term})\ \ \ We start 
from the Yukawa-type Klein Gordon equation as a basic wave equation\cite{rf2}.
\begin{eqnarray}
[  \partial^2/\partial X_\mu^2   &-& 
 {\cal M}^2 ( r_\mu ,\partial /\partial r_\mu ;
\partial /i \partial X_\mu  ) ] \Phi (X,r,\cdots ) =0\ ,\ \ 
 {\cal M}^2 = {\cal M}^2_{\rm conf} + \delta {\cal M}^2 ,\ \ \ \ \ \ \ \ 
\ \ \ \ \ 
\label{eq3}
\end{eqnarray}
where $X_\mu (r$ for mesons, $r_1,r_2$ for baryons) are the center of 
mass (relative) coordinates of hadron systems. In the squared mass 
operator ${\cal M}^2$ the confining-force part ${\cal M}^2_{\rm conf.}$ 
is assumed to be Lorentz-scalar and $A,(B)$-independent in the case of 
light-quark hadrons, leading to the mass spectra with the $\tilde U(12)$ 
symmetry and accordingly also with the chiral symmetry. As its concrete 
model we apply the covariant oscillator in COQM, leading to the 
straightly-rising Regge trajectories. The effects due to perturbative 
QCD and other possible effects $\delta {\cal M}^2$ are neglected in this 
paper. The WF are separated into the positive (negative)-frequency parts 
concerning the CM plane-wave motion and expanded in terms of eigen-states 
of the squared-mass operator, $\psi_N^{(\pm )}$ satisfying 
${\cal M}^2 \psi_N^{(\pm )} = M_N^2 \psi_N^{(\pm )}$, as   
\begin{eqnarray}
\Phi (X,r,\cdots) &=& \sum_N \sum_{{\bf P}_N}   
\left[  e^{iP_N\cdot X} \psi_N^{(+)} (P_N,r,\cdots )
     +e^{-iP_N\cdot X} \psi_N^{(-)} (P_N,r,\cdots ) \right] \ .\ \ \ \ \ 
\end{eqnarray}

\hspace*{-0.8cm}(\underline{\it BW-spinors\ and Covariant\ Oscillators\ 
as\ Expansion-bases})\ \ \ \ We will describe the internal WF of 
relativistic composite hadrons with a definite mass and a definite total 
spin $(J=L+S)$ which are tensors in the $\tilde U_{\rm DS}(4)\times O(3,1)$ 
space by expanding them in terms of covariant bases of complete set, 
being a direct product of eigen-functions, in the respective sub-space.
We choose the BW spinors and the covariant oscillator functions (with a 
definite metric type) as them.\\
(\underline{\it Spinor\ WF})\ \ \ \ The internal WF is, concerning the 
spinor freedom, expanded in terms of complete set of relevant multi-spinors, 
Bargmann-Wigner (BW) spinors. The BW spinors are defined as multi-Dirac 
spinor solutions of the relevant local Klein-Gordon equation:
\begin{eqnarray}
(  \partial^2/\partial X_\mu^2 &-& M^2  ) 
  W_{\alpha\cdots}^{\ \ \beta\cdots} (X) = 0
\label{eq9}\\  
W_{\alpha\cdots}^{\ \ \beta\cdots} (X) & \equiv & 
\sum_{{\bf P}}
(e^{iPX}W_{\alpha\cdots}^{(+)\beta\cdots}(P)
+ e^{-iPX}W_{\alpha\cdots}^{(-)\beta\cdots}(P)  ) .               
\end{eqnarray}
First we define the corresponding various free Dirac spinors for constituent 
quarks and anti-quarks with hadron 4-momentum $P_\mu$ as single-index 
BW spinors:
\begin{eqnarray}
\psi_{q,\alpha} (X)  &\equiv& W_\alpha (X)\ ,\ 
    u_{q,\alpha}(P_\mu )=W_\alpha^{(+)}(P)\ ,\  
  u_{q,\alpha}(-P_\mu )=W_\alpha^{(-)}(P)\ ,\  \ \ \  \label{eq211}\\
\psi_{\bar q,\alpha} (X)  &\equiv&  W_{\bar q,\alpha} (X)\ ,\  
    v_{\bar q,\alpha}(P_\mu )=W_\alpha^{(-)}(P)\ ,\  
  v_{\bar q,\alpha}(-P_\mu )=W_\alpha^{(+)}(P)\ .\  \ \ \  \label{eq212}
\end{eqnarray}
For mesons(baryons) the BW spinors are bi-Dirac (tri-Dirac) spinors.
\begin{eqnarray}
{\rm meson}: &\  & \psi^{(\pm )}_{N,A}{}^B (P_N,r) = \sum_W 
W_\alpha^{(\pm )\beta }(P_N) M^{(\pm ) b}_{N,a} (r,P_N)\\
{\rm baryon}: &\  & \psi^{(\pm )}_{N,A_1A_2A_3} (P_N,r_1,r_2) = \sum_W 
W_{\alpha_1\alpha_2\alpha_3}^{(\pm )}(P_N) 
B^{(\pm )}_{N,a_1a_2a_3}(r_1,r_2,P_N) .\ \ \ 
\end{eqnarray}
The irreducible composite hadrons are summarized in Tables \ref{tab6} 
and \ref{tab8}, respectively, for $q\bar q$-mesons and $qqq$-baryons. 
Here it is to be noted that the intrinsic spin-freedom for totality 
of BW spinors of $q\bar q$ mesons is $4\times 4=16$, four times of 
$2\times 2=4$ for boosted Pauli spinors. Correspondingly new types 
of BW spinors appear. For $qqq$ baryons $G(P)$ and $F(P)$, which 
include 1 and 2 negative energy Dirac components, respectively, 
appear in addition to the conventional $E(P)$ (with all positive 
energy Dirac components), boosted multi-Pauli spinor; and also to 
be noted that, although the BW equation with a definite mass itself 
is not $\tilde U_{\rm DS}(4)$ symmetric, the Klein-Gordon equation 
with a Lorentz-scalar mass-squared is generally invariant.\\
({\it Space-time\ WF})\ \ \ \ The internal WF is, concerning the 
relative space-time freedom, expanded in terms of the complete set 
of covariant oscillator eigen-functions, where (by applying a 
Lorentz-invariant subsidiary condition\cite{Takabayashi64} to 
``freeze" the relative-time freedom) the general symmetry $O(3,1)$ 
of the original ${\cal M}^2_{\rm conf}$ is reduced into the 
non-relativistic $O(3)$ symmetry.

Here it is noteworthy that the above choice of bases is desirable 
from the phenomenological facts  
i) that the constituent quark inside of hadrons behaves like a free 
Dirac-particle\footnote{
The BW-spinors with total hadron momentum $P_\mu$ and $M$ are easily
shown to be equivalent to the product of free Dirac spinors of the 
respective constituent ``exciton-quarks" with momentum 
$p_{N,\mu}^{(i)}\equiv \kappa^{(i)}P_{N,\mu}$ and mass 
$m_N^{(i)}\equiv \kappa^{(i)}M_N\ \ (\sum_i \kappa^{(i)}=1)$.
} (implied by BW-spinors) and  
ii) that in the global structure of hadron spectra (Regge trajectory 
and so on) is well described by the corresponding oscillator potential. \\
(\underline{\it Transformation\ rule\ for\ hadrons\ and\ chiral\ 
symmetry})\ \ \ \ By using the covariant quark representation of 
composite hadrons given above we can derive automatically their 
rule for any (relativistic) symmetry transformation from that of 
constituent quarks. The physical meaning of chiral transformation 
is clearly seen from the operations of its infinitesimal generator 
on the respective constituent quark spinors:\ 
$
u(P) \longrightarrow  u'(P)=-\gamma_5 u(P) =u(-P);
$\ \ 
$
v(P) \longrightarrow  v'(P)=-\gamma_5 v(P) =v(-P).
$
That is, the chiral transformation transforms the members of relevant 
BW-spinors with each other. Accordingly, if ${\cal M}^2$ operator is 
a Lorentz-scalar and independent of Dirac indices, the hadron mass 
spectra have effectively the $\tilde U(4)$ symmetry and accordingly, 
also the chiral symmetry. Here it is to be stressed that this is simply 
a phenomenological assumption. An intention is in this work not to treat 
a dynamical problem from a conventional composite picture, but is to 
propose a \underline{kinematical\ framework} for describing composite 
hadrons covariantly. The validity of the above assumption is checked 
only by comparing its predictions with experimental and phenomenological
facts.

\section{Level structure of mesons and baryons}

\hspace*{-0.8cm}(\underline{\it Assignment\ of\ mesons\ and\ baryons\ into\ 
$\tilde U_{\rm SF}(12)\bigotimes O(3,1)$\ scheme})\ \ \ \ We assign the 
light-quark ground state mesons and baryons to the representations
$(\underline{\bf 12}\times \underline{\bf 12}^*)=\underline{\bf 144}$ and 
$(\underline{\bf 12}\times\underline{\bf 12}\times 
\underline{\bf 12} )_{\rm Symm}=\underline{\bf 364}$, respectively.

In the extended version (old version) of COQM the confining force is 
assumed to be Lorentz-scalar (``boosted-spin" independent) and the 
mass spectra have the $\tilde U(12)_{\rm SF}$ symmetry (boosted 
$SU(6)_{\rm SF}$ symmetry). Thus, the WF of hadrons in the new 
level-classification scheme become the tensors in the 
$O(3,1)_{\rm Lorentz}\bigotimes \tilde U(4)_{\rm D.S.}\bigotimes$ 
$SU(3)_{\rm F}$-space(, being extended from the ones in the 
$O(3)\bigotimes SU(2)_{\rm P.S.}\bigotimes SU(3)_{\rm F}$ space of 
NRQM). The numbers of freedom of spin-flavor WF in NRQM are 
${\bf 6}\times {\bf 6}^*=\underline{\bf 36}$ for mesons and 
$({\bf 6}\times {\bf 6}\times {\bf 6})_{\rm Symm.}
=\underline{\bf 56}$ for baryons: These numbers in COQM become 
${\bf 12}\times {\bf 12}^*=\underline{\bf 144}$ for mesons and 
 $({\bf 12}\times {\bf 12}\times {\bf 12})_{\rm Symm.}=
\underline{\bf 364}=\underline{\bf 182}$ (for baryons)
$+\underline{\bf 182}$ (for anti-baryons).

Inclusion of heavy quarks is straightforward: The WF of general 
$q$ and$/$or $Q$ hadrons become tensors in $O(3,1)\bigotimes 
[\tilde U(4)_{\rm D.S.}\bigotimes SU(3)_{\rm F}]_q
\bigotimes [SU(2)_{\rm P.S.}\bigotimes U(1)_{\rm F}]_Q$.\\
(\underline{\it Level\ structure\ of\ ground\ state\ 
mesons})\cite{IIM00}\ \ \ \ In Table \ref{tab6} we have summarized 
the properties of ground state mesons in the light and$/$or heavy 
quark systems. It is remarkable that there appear new multiplets 
of the scalar and axial-vector mesons in the $q$-$\bar Q$ and 
$Q$-$\bar q$ systems and that in the $q$-$\bar q$ systems the 
two sets (Normal and Extra) of pseudo-scalar and of vector meson 
nonets exist. The $\pi$ nonet ($\rho$ nonet) is assigned to the
$P_s^{(N)}\ (V_\mu^{(N)})$ state. We call the new type of particles 
in the extended COQM (which have never appeared in NRQM) as 
``chiralons''.

\begin{table}
\begin{center}
\begin{tabular}{l|c|l|lcl}
\hline
   & mass & Approx. Symm. & Spin WF & $SU(3)$ & Meson Type \\
\hline
$Q\bar Q$ & $m_Q+m_{\bar Q}$ & $LS$ symm. & $u_Q(P)\bar v^{\bar Q}(P)$
          & $\underline{1}$ & $P_s,\ V_\mu$  \\
\hline
$q\bar Q$ & $m_q+m_{\bar Q}$ & $q$-Chiral Symm. & $u_q(P)\bar v^{\bar Q}(P)$
          & $\underline{3}$ & $P_s,\ V_\mu$  \\
               &                              & $\bar Q$-Heavy Q. Symm. 
          & $u_q(-P)\bar v^{\bar Q}(P)$ & $\underline{3}$ & $S,\ A_\mu$  \\
$Q\bar q$ & $m_Q+m_{\bar q}$ & $\bar q$-Chiral Symm.
          & $u_Q(P)\bar v^{\bar q}(P)$
          & $\underline{3}^*$ & $P_s,\ V_\mu$  \\
               &                              & $Q$-Heavy Q. Symm. 
          & $u_q(P)\bar v^{\bar Q}(-P)$ & $\underline{3}^*$ & $S,\ A_\mu$  \\
\hline
$q\bar q$ & $m_q+m_{\bar q}$ & Chiral Symm.
     & $\frac{1}{\sqrt{2}}(u(P)\bar v(P)\pm u(-P)\bar v(-P) )$
     & \underline{\bf 9} & $P_s^{(N,E)},\ V_\mu^{(N,E)}$  \\
              &                             &   
     & $\frac{1}{\sqrt{2}}(u(P)\bar v(-P)\pm u(-P)\bar v(P) )$
     & \underline{\bf 9}   &  $S^{(N,E)},\ A_\mu^{(N,E)}$  \\
\hline
\end{tabular}
\end{center}
\caption{Level structure of ground-state mesons}
\label{tab6}
\end{table}

\hspace*{-0.8cm}(\underline{Level\ structure\ of\ mesons\ 
in\ general})\cite{IIM00}\ \ \ \ The global mass spectra of the 
ground and excited state mesons are given by
\begin{eqnarray} 
M_N^2 & = & M_0^2+N\Omega = m_N^{(1)} + m_N^{(2)} .
\label{eq41}
\end{eqnarray}
Their quantum numbers are given in Table \ref{tab7}. Here it is 
to be noted that some chiralons have the ``exotic" quantum numbers
from the conventional NRQM viewpoint.

\begin{table}
\begin{center}
\begin{tabular}{c@{}||@{}c} \hline
$\begin{array}{c|c|c|c}
(q\bar q)  & P & C & N \\
\hline
\{P_s^{(N)},V_\mu^{(N)}  \} \bigotimes \{ L,N \} & 
(-1)^{L+1} & (-1)^{L+S} & {\rm all}\\
\{P_s^{(E)},V_\mu^{(E)}  \} \bigotimes \{ L,N \} & 
(-1)^{L+1} & (-1)^{L+S} & 0,1\\
\end{array}$
 &  
$\begin{array}{c|c|c}
 (q\bar Q\  {\rm or}\  Q\bar q) & P & N \\
\hline
 \{ P_s,V_\mu \} \bigotimes \{ L,N \} & (-1)^{L+1} & {\rm all}\\ 
 \{ S,A_\mu \} \bigotimes \{ L,N \} & (-1)^{L} & 0,1\\
\hline
\end{array}$ \\
$\begin{array}{c|c|c|c}
\{ S^{(N)},A_\mu^{(N)}  \} \bigotimes \{ L,N \} & 
(-1)^{L\ \ \ } & (-1)^{L} & 0,1\\
\{ S^{(E)},A_\mu^{(E)}  \} \bigotimes \{ L,N \} & 
(-1)^{L\ \ } & (-1)^{L+1} & 0,1\\
\end{array}$ 
 & 
$\begin{array}{c|c|c}\hline 
 (Q\bar Q) & P & N \\
\hline
 \{ P_s,V_\mu \} \bigotimes \{ L,N \} & (-1)^{L+1} & {\rm all} \\
\end{array}$ \\
\hline
\end{tabular}
\end{center}
\caption{Level structure of Mesons in general:
We are able to infer\cite{protovshin} that the chiral symmetry concerning the light quarks is valid (still effective)
for the ground (first excited) state of $n\bar n$ and $n\bar Q$ meson systems, while the
symmetry will prove invalid from the $N$-th ($N\geq 2$) excited hadrons.}
\label{tab7}
\end{table}

\hspace*{-0.8cm}(\underline{Level\ Structure\ of\ Baryons})\ \ \ \  
The baryon WF in Eq.~(\ref{eq2}) should be full-symmetric (except 
for the color freedom) under exchange of constituent quarks: The 
full-symmetric total WF in the extended scheme is obtained, in 
the following three ways, as a product of the sub-space WF with
respective symmetric properties:
\begin{eqnarray}	
| \rho F \sigma \rangle_S & = & | \rho \rangle_S |F \sigma\rangle_S\ ;
\ \ \frac{1}{\sqrt{2}}(
 | \rho \rangle_\alpha |F \sigma\rangle_\alpha 
   + | \rho \rangle_\beta |F \sigma\rangle_\beta );\ \ 
 | F \rangle_A |\rho\sigma\rangle_A\  ;\ \ \ \ \ 
\end{eqnarray}
where $\rho\bigotimes\sigma =\gamma$ is the conventional two, $\rho$ 
and $\sigma$ spin, 2 by 2 matrix representation of the 4 by 4 Dirac 
matrix. $|\ \ \rangle_S$, $|\ \  \rangle_{\alpha (\beta )}$ and 
$|\ \ \rangle_A$ mean the full-symmetric, $\alpha (\beta)$-type partial 
symmetric and full anti-symmetric subspace WF, respectively. The 
intrinsic parity operation is given by $\hat P = \Pi_{i=1}^3 
\gamma_4^{(i)}$, that is, the parity of $(E^{(+)},G^{(+)},F^{(+)})$ 
BW spinors are $(+,-,+)$ and those of $(E^{(-)},G^{(-)},F^{(-)})$ 
BW spinors are $(-,+,-)$. The symmetry properties of ground state 
light-quark baryon WF and their level structures thus determined 
are summarized in Table \ref{tab8}.  

\begin{table}
\begin{center}
\begin{tabular}{clcrc}
\hline
$W^{(+)}$ & spin-flavor wave function & $B^{P\hspace{-0.22cm}\bigcirc}$ 
  &  static & $SU(6)$ \\ 
\hline
$E^{(+)}$ : & 
$|\rho\rangle_S |F\sigma\rangle_S=|\rho\rangle_S |F\rangle_S|\sigma\rangle_S$ 
 & $\Delta_{3/2}^{+\hspace{-0.22cm}\bigcirc}$ & $10\times 4=40$ &   \\
  & \ \ \ \ \ \ \ \ \ \ \ \ \ \ \ \ \ 
$|\rho\rangle_S \frac{1}{\sqrt{2}}(|F\rangle_\alpha |\sigma\rangle_\alpha +
|F\rangle_\beta |\sigma\rangle_\beta )$ 
 & $N_{1/2}^{+\hspace{-0.22cm}\bigcirc}$ & $8\times 2=16$ & \underline{\bf 56} \\
\hline
$G^{(+)}$ : & 
$\frac{1}{\sqrt{2}}(|\rho\rangle_\alpha |F\sigma\rangle_\alpha 
       +|\rho\rangle_\beta |F\sigma\rangle_\beta )$ ; 
$|F\sigma \rangle_{\alpha (\beta )} = |F\rangle_S |\sigma\rangle_{\alpha (\beta )}$
 & $\Delta_{1/2}^{-\hspace{-0.22cm}\bigcirc}$ & $10\times 2=20$ &   \\
  & \ \ \ \ \ \ \ \ \ \ \ \ \ \ 
\ \ \ \ \ \ \ \ \ \ \ \ \ \ \ \ \ \ \ \ \ \ \ \ \ \ \ \ \ \ \ \ \ \ 
$|F\rangle_{\alpha (\beta )} |\sigma\rangle_S$ 
 & $N_{3/2}^{-\hspace{-0.22cm}\bigcirc}$ & $8\times 4=32$ &    \\
  & \ \ \ \ \ \ \ \ 
    $| F \sigma \rangle_{\alpha (\beta )} = \frac{1}{\sqrt{2}}
         ( \mp |F\rangle_\alpha |\sigma\rangle_{\alpha (\beta )} 
           +|F\rangle_\beta |\sigma\rangle_{\beta (\alpha )})$ 
 & $N_{1/2}^{-\hspace{-0.22cm}\bigcirc}$ & $8\times 2=16$ &    \\
 & 
$|F\rangle_A  |\rho\sigma\rangle_A=|F\rangle_A
\frac{1}{\sqrt{2}}(-|\rho\rangle_\alpha |\sigma\rangle_\beta
   +|\rho\rangle_\beta |\sigma\rangle_\alpha )$  
 & $\Lambda_{1/2}^{-\hspace{-0.22cm}\bigcirc}$ & $1\times 2=2$ & 
 \underline{\bf 70} \\
\hline
$F^{(+)}$ : & 
$|\rho\rangle_S |F\sigma\rangle_S=|\rho\rangle_S |F\rangle_S|\sigma\rangle_S$ 
 & $\Delta_{3/2}^{+\hspace{-0.22cm}\bigcirc}$ & $10\times 4=40$ &   \\
  & \ \ \ \ \ \ \ \ \ \ \ \ \ \ \ \ \ 
$|\rho\rangle_S \frac{1}{\sqrt{2}}(|F\rangle_\alpha |\sigma\rangle_\alpha +
|F\rangle_\beta |\sigma\rangle_\beta )$ 
 & $N_{1/2}^{+\hspace{-0.22cm}\bigcirc}$
 & $8\times 2=16$ & $\underline{\bf 56}^\prime$ \\
\hline
\end{tabular}
\end{center}
\caption{Level structure of ground-state $qqq$-baryon:
 $_{12}H_3\ /2=\underline{\bf 364}\ /2=
\underline{\bf 182}$.
}
\label{tab8}
\end{table}
 
Here it is remarkable that there appear chiralons in the ground states.
That is, the extra positive parity $\underline{\bf 56}^\prime$-multiplet of the 
static $SU(6)$ and the extra negative parity \underline{\bf 70}-multiplet 
of the $SU(6)$ in the low mass region. It is also to be noted that the 
chiralons in the first excited states are expected to exist. The above 
consideration on the light-quark baryons are extended directly to the 
general light and$/$or heavy quark baryon systems: The chiralons are 
expected to exist also in the $qqQ$ and $qQQ$-baryons, while no chiralons 
in the $QQQ$ system.

\section{Experimental Candidates for Chiral Particles}

In our level-classification scheme a series of new type of multiplets 
of the particles, chiralons, are predicted to exist in the ground and 
the first excited states of $q\bar q$ and $q\bar Q$ or $Q\bar q$ meson 
systems and of $qqq$, $qqQ$ and $qQQ$-baryon systems. Presently we can 
give only a few experimental candidates or indications for them:\\
\underline{\em ($q\bar q$-mesons)}\ \ \ \ One of the most important 
candidates is the scalar $\sigma$ nonet to be assigned as $S^{(N)}(^1S_0)$ : 
$[\sigma (600),\ \kappa (900),\ a_0(980),\ f_0(980)]$. The existence of 
$\sigma (600)$ seems to be established\cite{rf1} through the analyses of, 
especially, $\pi\pi$-production processes. A firm experimental 
evidence\cite{rfP3} for $\kappa (800$--$900)$ through the decay 
process\cite{rfP4} $D^+\rightarrow K^-\pi^+\pi^+$ was reported recently 
at the conference, Hadron2001.\\
In our scheme respective two sets of $P_s$- and of $V_\mu$-nonets, to 
be assigned as $P_s^{(N,E)}(^1S_0)$ and $V_\mu^{(N,E)}(^3S_1)$, are 
to exist: Out of the five vector mesons (stressed\cite{rfP5} as problems 
with vector mesons, $[\rho^\prime (1450),\  \rho^\prime (1700),\ 
\omega^\prime (1420),\  \omega^\prime (1600),\  \phi (1690) ]$, the 
lower mass $\rho^\prime (1450)$ and $\omega^\prime (1420)$, and the 
$\phi (1690)$ are naturally able to be assigned as the members of 
$V_\mu^{(E)}$-nonets;\\
Out of the three established $\eta$, $[\eta (1295),\ \eta (1420),\  
\eta (1460) ]$ at least one extra, plausibly $\eta (1295)$ with the 
lowest mass, may belong to $P_s^{(E)}(^1S_0)$ nonet.\\
Recently the existence of two ``exotic" particles $\pi_1(1400)$ and 
$\pi_1(1600)$ with $J^{PC}=1^{-+}$ and $I=1$, observed\cite{rfP6}   
in the $\pi\eta ,\  \rho\pi$ and other channels, is attracting strong 
interests among us.
These exotic particles with a mass around 1.5GeV may be naturally 
assigned as the first excited states $S^{(E)}(^1P_1)$ and 
$A_\mu^{(E)}(^3P_1)$ of the chiralons.\\
(\underline{$q\bar Q$}\ or\ \underline{$Q\bar q$}-mesons)\ \ \ \ Recently 
we have shown at the conference some indications for existence of the 
two chiralons in $D$- and $B$-meson systems (\cite{rfP7} and \cite{rfP8}) 
obtained through analyses of the $\Upsilon (4S)$ or $Z^0$ decay processes, 
respectively,\\
\hspace*{2cm}
$D_1^\chi = A_\mu (^3S_1), \ \  J^P=1^+\ \ 
{\rm in}\ \ D_1^\chi \rightarrow D^* + \pi$, \\
\hspace*{2cm}
$B_0^\chi = S (^1S_0), \ \  J^P=0^+\ \ 
{\rm in}\ \ B_0^\chi \rightarrow B + \pi  $.\\   
(\underline{\em $qqq$}-baryons)\ \ \ \  The two facts have been a 
longstanding problem that the Roper resonance $N(1440)_{1/2^+}$ 
is too light to be assigned as radial excitation of $N(939)$ and 
that $\Lambda (1405)_{1/2^-}$ is too light as the $L=1$ excited 
state of $\Lambda (1116)$. In our new scheme they are reasonably 
assigned to the members of ground state chiralons with 
$[SU(6),\ SU(3),\ J^P]$, respectively, as \\
\hspace*{1cm}
$N(1440)_{1/2^+} = F(\underline{\bf 56}^\prime ,\ \underline{\bf 8},\  {1}/{2}^+),\ \ 
\Lambda (1405)_{1/2^-} = G(\underline{\bf 70},\ \underline{\bf 1},\   {1}/{2}^-)$\ .\\
The particle $\Delta (1600)_{3/2^+}$ which is lighter than 
$\Delta (1620)_{1/2^-}$ may also belong to the extra 
$\underline{\bf 56}^\prime$ of the ground state chiralons. 
This situation is shown in Table \ref{tab9}. 

\begin{table}
\begin{tabular}{c|cl|cl}
\hline
SU(6)  & SU(3), $J^P$ &  &  SU(3), $J^P$ &   \\
\hline
\underline{\bf 56} & \underline{\bf 8},\ $\frac{1}{2}^+$ 
     & $N(939),\Lambda (1116),\Sigma (1192),
\Xi (1318)$ & \underline{\bf 10},\ $\frac{3}{2}^+$
     & $\Delta (1232),\Sigma (1385),\cdots $ \\
\underline{\bf 56}$^\prime$ & \underline{\bf 8},\ $\frac{1}{2}^+$ & 
$\begin{array}{|c|}\hline N(1440)\\  \hline \end{array}$,\ \ \ \ \ \ \ 
$\begin{array}{|c|}\hline \Sigma (1660)\\  \hline \end{array}$ 
  & \underline{\bf 10},\ $\frac{3}{2}^+$ & 
$\begin{array}{|c|}\hline \Delta (1600)  \\  \hline \end{array}$   \\
\hline 
\underline{\bf 70} & \underline{\bf 8},\ $\frac{1}{2}^-$ & 
$N(1535)$  & \underline{\bf 10},\ $\frac{1}{2}^-$ & $\Delta (1620)$   \\
   & \underline{\bf 1},\ $\frac{1}{2}^-$ & 
---------$\begin{array}{|c|}\hline \Lambda (1405)\\  \hline \end{array}$-----------------  &  &   \\
\hline
\end{tabular}
\caption{Assignment of $qqq$-baryons: The baryons in the boxes are candidates of chiralons.}
\label{tab9}
\end{table}

\section{Concluding Remarks}

We have presented an attempt for Level-classification scheme unifying
the seemingly contradictory two viewpoints, Non-relativistic one with 
$LS$-symmetry and Relativistic one with Chiral symmetry . As results, 
We have predicted the existence of \underline{New Chiral Particles 
in the lower mass regions} ``Chiralons'', which had never been appeared 
in NRQM. We have several good candidates for chiralons, for example,\\
$\sigma$-nonet\ \  \{  
                 $\sigma (600)$, $\kappa (800)$, $a_0(980)$, $f_0(980)$   \}
                {\em as ``Relativistic'' S-wave states of $q\bar q$}; 
$\pi_1(1400)$ and $\pi_1 (1600)$ with $J^{PC}=1^{-+}$ 
                {\em as ``Relativistic'' P-wave states of $q\bar q$};  
Roper resonance $N(1440)_{1/2^+}$ and 
             SU(3) singlet $\Lambda (1405)_{1/2^-}$
     {\em as ``Relativistic'' S-wave states of $qqq$}.   

Further search for chiralons, both experimental and theoretical, 
is urgently required for developing hadron physics.

%

\end{document}